\documentclass[11pt,twocolumn,english]{article}
\usepackage[latin9]{inputenc}
\usepackage{babel}
\usepackage{amsmath}
\usepackage{graphicx}
\usepackage{esint}
\usepackage[numbers]{natbib}
\usepackage[unicode=true]
 {hyperref}

\pdfpageheight\paperheight
\pdfpagewidth\paperwidth

\newcommand{\noun}[1]{\textsc{#1}}

\usepackage{times}
\usepackage{calc}
\begin{document}
\title{Deep reinforced learning heuristic tested on spin-glass ground states:
The larger picture}
\author{Stefan Boettcher\thanks{sboettc@emory.edu}, Department of Physics,
Emory University, Atlanta, GA 30322; USA}
\date{}
\maketitle
\noun{Matters Arising} from Changjun Fan et al. \emph{Nature Communications}
https://doi.org/10.1038/s41467-023-36363-w (2023).

\textbf{In Ref.~\cite{Fan23}, the authors present a deep reinforced
learning approach to augment combinatorial optimization heuristics.
In particular, they present results for several spin glass ground
state problems~\cite{Dagstuhl04}, for which instances on non-planar
networks are generally NP-hard, in comparison with several Monte Carlo
based methods, such as simulated annealing (SA) or parallel tempering
(PT)~\cite{Wang15}. Here, we examine those results in the context
of well-established literature and find that, albeit fast and capable
for small instance sizes, the presentation lacks signs of the claimed
superiority for larger instances, unless one competes with Greedy
Search for speed.}

Indeed, the results of Ref.~\cite{Fan23} demonstrate that the reinforced
learning improves the results over those obtained with SA or PT, or
at least allows for reduced runtimes for the heuristics before results
of comparable quality have been obtained relative to those other methods.
To facilitate the conclusion that their method is ``superior'',
the authors of Ref.~\cite{Fan23} pursue two basic strategies: (1)
A commercial GUROBI solver (see \href{https://www.gurobi.com/}{https://www.gurobi.com/})
is called on to procure a sample of exact ground states as a testbed
to compare with, and (2) a head-to-head comparison between the heuristics
is given for a sample of larger instances where exact ground states
are hard to ascertain. Here, we put these studies into a larger context,
showing that the claimed superiority is at best marginal for smaller
samples and becomes essentially irrelevant with respect to any sensible
approximation of true ground states in the larger samples. For example,
this method becomes irrelevant as a means to determine stiffness exponents
$\theta$ in $d>2$, as mentioned by the authors, where the problem
is not only NP-hard but requires the subtraction of two almost equal
ground-state energies and systemic errors in each of $\approx1\%$
found here are unacceptable~\cite{Boettcher05d}. This larger picture
on the method arises from a straightforward finite-size corrections
study over the spin glass ensembles the authors employ, using data
that has been available for decades~\cite{Pal96,EOSK}.

In our investigation here, we focus on mainly two ensembles of NP-hard
problems the authors utilize: The Edwards-Anderson spin glass on a
cubic lattice (EA in $d=3$) with periodic boundary conditions~\cite{Edwards75}
and the mean-field (all-to-all connected) Sherrington-Kirkpatrick
spin glass (SK)~\cite{Sherrington75}. The ensemble for both models
consists of instances where all bonds are chosen randomly from a normal
distribution of zero mean and unit variance. The ensemble is parametrized
by its size, i.e., the number of variables $N$ in a spin configuration
$\vec{\sigma}$, where $N=L^{3}$ in the case of EA. With those hard
combinatorial problems, there are many ways to find exact solutions
for instances of small $N$, such as a solver like GUROBI, however,
for any practical application at large $N$, the super-polynomial
rise in complexity necessitates the use of heuristic methods. Thus,
the scalability of a heuristic is of particular concern. In the formal
study of computational complexity, this is typically addressed by
establishing bounds on an all-encompassing worst-case scenario~\cite{G+J}.
For many complicated meta-heuristics~\cite{Voss99}, such as the
case of the method presented here, insights into the capability of
a heuristic can be gained only from comparative studies over widely
accepted testbeds of instances or those selected from specific ensembles.
The authors have clearly adopted the ensemble approach~\cite{Fan23}.

Especially with regard to scaleability, the ensemble picture deserves
particular attention, for the following reasons. Those ensembles typically
have a ``thermodynamic limit'', i.e., their averages are well-defined
and possess a clear meaning for $N\to\infty$, which a typical large
instance approaches. At times, that limit may even be exactly solvable,
such as in the case of SK~\cite{MPV}, but that is not essential
here, as exemplified by EA. More importantly, that limit is usually
attained in an equally well-defined manner through finite-size corrections
(FSC). To be specific in this context, for the cost function a heuristic
is trying to minimize, the authors have chosen the ground state energy
density, $e_{0}=\min_{\vec{\sigma}}H\left(\vec{\sigma}\right)/N$,
of the Hamiltonian $H$ for each of their (physically motivated) spin
glass ensembles. Instances are generated via random choices of bonds
$J_{ij}$ from a characteristic distribution $P(J)$, see Eq.~(1)
in Ref.~\cite{Fan23}. If the thermodynamic limit for the ensemble-averaged
ground-state energy density $\left\langle e_{0}\right\rangle _{N=\infty}$
exists, FSC assumes the asymptotic scaling form 
\begin{equation}
\left\langle e_{0}\right\rangle _{N}\sim\left\langle e_{0}\right\rangle _{N=\infty}+\frac{A}{N^{\omega}}+\ldots,\qquad(N\to\infty),\label{eq:FSC}
\end{equation}
for a constant $A$ and a correction exponent $\omega(>0)$. Clearly,
other forms of corrections might exist and higher-order terms could
well obscure the assumed behavior deep into the large-$N$ regime.
Yet, self-consistency with the form in Eq.~(\ref{eq:FSC}) of the
actual data for small $N$, where reliable (or exact) results can
be ascertained, often provides a powerful baseline to assess the scalability
of a heuristic~\cite{BoFa11,Boettcher19}. This is certainly the
case here, and it provides a larger picture for the results in Ref.~\cite{Fan23}.

\begin{figure}
\vspace{-0.2cm}

\hfill{}\includegraphics[clip,width=1\columnwidth,viewport=0bp 10bp 740bp 550bp]{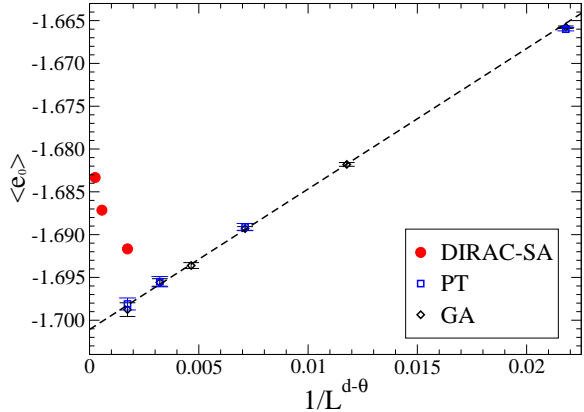}\hfill{}

\vspace{-0.3cm}

\caption{\label{fig:EA3dFSC}Extrapolation plot according to the finite-size
corrections form in Eq.~(\ref{eq:FSC}) for the ensemble averaged
ground state energy densities obtained with various heuristics for
EA in $d=3$. Previous data obtained with GA~\cite{Pal96} or PT~\cite{Wang15}
for a range of system sizes $N=L^{d}$ up to $L=14$ exhibit a consistent
asymptotic scaling with corrections $\sim1/N^{\omega}$ and $\omega=1-\theta/d\approx0.92$,
as discussed in Ref.~\cite{BoFa11}. The linear fit (dashed line)
with $x=1/L^{2.76}$ has the form $\left\langle e_{0}\right\rangle _{N=\infty}+Ax$
with $\left\langle e_{0}\right\rangle _{N=\infty}\approx-1.701$ and
$A\approx1.641$. The corresponding data for $L=10,15,20$ from Ref.~\cite{Fan23}
(red circles) diverges increasingly from the expected values for typical
ground states.}
\end{figure}

Long before the PT results~\cite{Wang15} that the authors reference
in their study of EA in $d=3$, virtually identical results have been
found by Pal~\cite{Pal96} using a genetic algorithm (GA). Despite
the doubts the authors raise (in the caption \footnote{Note that several references in Ref.~\cite{Fan23} are incorrect,
e.g., in the caption to Fig.~5 ``Ref.~51'' should be to Ref. 50
and the label ``(f)'' should be~(d) for the \emph{3d}-EA at $L=10$.} of their Fig.~5), both the PT and the GA data exhibit a consistent
scaling picture, shown here in Fig.~\ref{fig:EA3dFSC}. While the
authors don't provide any tabulated data for their corresponding results,
at least for the larger samples we can extract estimated values for
their best results (for DIRAC-SA, shown as red circles in Fig.~\ref{fig:EA3dFSC})
from the plots provided in their Fig.~S5 (d-f). There, the fact that
the DIRAC-SA data is better than either PT or SA is taken as evidence
of the superiority of their method by the authors. However, considering
how far separated from any actual ground states every one of the datasets
employed in this comparison really is, this advantage, whether in
speed or in accuracy, is rather inconsequential in the larger picture
of Fig.~\ref{fig:EA3dFSC}.

\begin{figure}
\hfill{}\includegraphics[clip,width=1\columnwidth,viewport=10bp 130bp 590bp 520bp]{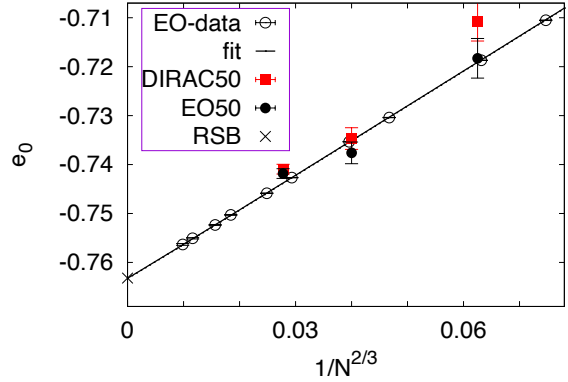}\hfill{}

\caption{\label{fig:SKFSC} Extrapolation plot of ensemble-averaged ground
state energy densities for SK according to Eq.~(\ref{eq:FSC}) with
$\omega=2/3$. For SK, theory (RSB~\cite{MPV}) predicts an exact
result for the limit $N\to\infty$, $\left\langle e_{0}\right\rangle _{N=\infty}=-0.7632\ldots$,
marked by $\times$. The reference data (open circles) for up to $N=1023$,
averaged over at least $10^{5}$ instances each, was obtained with
the extremal optimization heuristic (EO)~\cite{EOSK}. That the
asymptotic fit (line) of this data predicts $\left\langle e_{0}\right\rangle _{N=\infty}$
with high accuracy adds confidence in the scaling. The data for 50
instances each at $N=64,$ 125, and 216 from Ref.~\cite{Fan23}
(red squares) matches within errors to a similar random sample of
50 instances each optimized with EO (filled circles). Note that the
ground-state energy variances for SK are typically broader than standard
deviations~\cite{EOSK}. The DIRAC50 data here was obtained from
Fig.~S15 (d-f) in Ref.~\cite{Fan23}, which required the addition
division by $\sqrt{N}$ in the ground state energy densities when
a univariate ($\left\langle J^{2}\right\rangle =1$) bond distribution
is used, i.e., $\left\langle e_{0}\right\rangle _{N=64}=-5.6856/\sqrt{64}=-0.7107$,
$\left\langle e_{0}\right\rangle _{N=125}=-8.2141/\sqrt{125}=-0.7347$,
and $\left\langle e_{0}\right\rangle _{N=216}=-10.8895/\sqrt{216}=-0.7409$.}
\end{figure}

Similarly, the results the authors provide for SK prove inconclusive
in the larger picture of long-established results for this case~\cite{EOSK,Aspelmeier07,Boettcher19}.
Here, Ref.~\cite{Fan23} merely provides results of their method
for quite small instances, where GUROBI allows to obtain exact ground
states for comparison. While these results are indeed consistent with
the predicted scaling, as shown in Fig.~\ref{fig:SKFSC}, the sizes
bounded by $N\leq216$ considered in their study have very limited
predictive power about the scalability of their method for any size
that would make their method competitive, either in speed or in accuracy,
with state-of-the-art heuristics at larger $N$. After all, with an
ensemble approach, it is not necessary to rely on exactly solved instances
to make impactful comparisons, as our discussion of EA demonstrates.

In conclusion, a comparison with existing data shows little evidence
for the claimed superiority of the deep reinforcement learning strategy
to enhance optimization heuristics proposed in Ref.~\cite{Fan23}.
The comparison provided here for both, a sparse short-range and a
dense infinite-range spin glass model, is quite exemplary for all
the ensembles the authors discuss, so that this conclusion is likely
not particular to these two cases. The authors should be lauded for
having demonstrated some gains relative to simple greedy algorithms
for EA~\cite{Boettcher22}, but their results remain too far from
optimality, even if under the $<1\%$ level we found in Fig.~\ref{fig:EA3dFSC},
to be of any use in applications to the physics of spin glasses the
authors imply. For example, in the stiffness problem one determines
a ground state of an instance in EA and again for reversed boundary
conditions, which inserts a relative domain wall between the ground
states with separate energies $e_{0}^{1,2}(L)\sim\left\langle e_{0}\right\rangle _{L=\infty}+A_{1,2}/L^{d\omega}+\ldots$.
That domain wall has a much smaller energy, $\Delta e=\left|e_{1}-e_{2}\right|\sim\Delta A/L^{d\omega}\to0$,
which relates FSC to the stiffness exponent via $d\omega=d-\theta$~\cite{BoFa11},
as used in Fig.~\ref{fig:EA3dFSC}. These exponents were determined
for EA in dimensions $d=3,\ldots,7$ by finding ground states for
millions of dilute lattices with up to $N=10^{7}$ using a hybrid
EO algorithm~\cite{Boettcher04c,Boettcher05d}. Hence, the heuristics
chosen as a base for their comparison is surprisingly narrow, considering
that the authors refer to Ref.~\cite{Dagstuhl04} for the use of
heuristics for spin glasses, which also discusses GA and EO.

\subsection*{Competing Interests}

The author declares no competing interests.

\subsection*{Author Contributions}

The author assembled the research presented in this manuscript and
wrote the paper.

\paragraph*{Correspondence}

and requests for materials should be addressed to the author.

\bibliographystyle{unsrtnat}
\bibliography{/Users/sboettc/Boettcher}

\end{document}